\documentclass[twocolumn,reprint,amsmath,amssymb,aps,pra,superscriptaddress,bibnotes,longbibliography]{revtex4-1}

\usepackage{times}

\usepackage{amsmath}
\usepackage{amsfonts}
\usepackage{amssymb}
\usepackage{graphicx}
\usepackage{hyperref}

\usepackage{amstext}
\usepackage{amsxtra}
\usepackage{float}
\usepackage{upgreek}

\usepackage{subfigure}
\usepackage{braket}

\usepackage{textcomp}
\usepackage{natbib}

\hypersetup{backref,
colorlinks=true,
linkcolor=cyan,
linktoc=black,
citecolor=cyan,
urlcolor=cyan}

\newcommand{\kB}{k_{\textrm{B}}}

\begin{document}

\title{From single-particle excitations to sound waves in a box-trapped atomic Bose-Einstein condensate}

\author{Samuel J. Garratt}
\affiliation{Cavendish Laboratory, University of Cambridge, J. J. Thomson Avenue, Cambridge CB3 0HE, United Kingdom}
\affiliation{Theoretical Physics, University of Oxford, Parks Road, Oxford OX1 3PU, United Kingdom}
\author{Christoph Eigen}
\affiliation{Cavendish Laboratory, University of Cambridge, J. J. Thomson Avenue, Cambridge CB3 0HE, United Kingdom}
\author{Jinyi Zhang}
\affiliation{Cavendish Laboratory, University of Cambridge, J. J. Thomson Avenue, Cambridge CB3 0HE, United Kingdom}
\author{Patrik Turz\'{a}k}
\affiliation{Cavendish Laboratory, University of Cambridge, J. J. Thomson Avenue, Cambridge CB3 0HE, United Kingdom}
\author{Raphael Lopes}
\affiliation{Cavendish Laboratory, University of Cambridge, J. J. Thomson Avenue, Cambridge CB3 0HE, United Kingdom}
\affiliation{Laboratoire Kastler Brossel, Coll{\`e}ge de France, CNRS, ENS-PSL University, UPMC-Sorbonne Universit{\'e}s, 11 Place Marcelin Berthelot, F-75005 Paris, France}
\author{Robert P. Smith}
\affiliation{Cavendish Laboratory, University of Cambridge, J. J. Thomson Avenue, Cambridge CB3 0HE, United Kingdom}
\affiliation{Clarendon Laboratory, University of Oxford, Parks Road, Oxford OX1 3PU, United Kingdom}
\author{Zoran Hadzibabic}
\affiliation{Cavendish Laboratory, University of Cambridge, J. J. Thomson Avenue, Cambridge CB3 0HE, United Kingdom}
\author{Nir Navon}
\affiliation{Department of Physics, Yale University, New Haven, Connecticut 06520, USA}

\begin{abstract}
We experimentally and theoretically investigate the lowest-lying axial excitation of an atomic Bose-Einstein condensate in a cylindrical box trap.
By tuning the atomic density, we observe how the nature of the mode changes from a single-particle excitation (in the low-density limit) to a sound wave (in the high-density limit). Throughout this crossover the measured mode frequency agrees with Bogoliubov theory. Using approximate low-energy models we show that the evolution of the mode frequency is directly related to the interaction-induced shape changes of the condensate and the excitation.  
Finally, if we create a large-amplitude excitation, and then let the system evolve freely, we observe that the mode amplitude decays non-exponentially in time; this nonlinear behaviour is indicative of interactions between the elementary excitations, but remains to be quantitatively understood.
\end{abstract}

\maketitle

\section{Introduction}
Low-energy excitations play a central role in our understanding of many-body systems.
They characterise a system's low-temperature thermal properties, its response to small perturbations, and its near-equilibrium transport behaviour.
The collective excitations of ultracold gases have been extensively studied in the traditional setting of a harmonic trap for bosons \cite{JinModes,MewesModes,andrews1997propagation,marago2000observation, Steinhauer,meppelink2009sound,lobser2015observation} and fermions \cite{bartenstein2004collective,joseph2007measurement,nascimbene2009collective,sidorenkov2013second} with contact interactions (including low-dimensional gases \cite{moritz2003exciting,vogt2012scale}), as well as for Bose-Fermi mixtures \cite{ferrier2014mixture}, spin-orbit coupled gases \cite{zhang2012collective}, and gases with dipolar interactions \cite{bismut2010collective,chomaz2018observation}.

The recent developments in creating quasi-uniform box traps \cite{GauntUniform,chomaz2015emergence,ZwierleinUniform,HueckUniform} have led to intriguing new possibilities.
These traps provide a textbook setting for the study of short-wavelength excitations \cite{RaphaBragg}, but they also raise new questions on the nature of long-wavelength (system-size) collective modes, as highlighted by recent studies of sound propagation in 3D Bose \cite{Turbulence} and Fermi \cite{MITfootnote} gases, and 2D Bose gases \cite{ville2018sound} (see also \cite{ota2018second,ota2018collisionless,cappellaro2018collisionless}).
Due to the hard-wall boundary conditions the dynamics depend only on the interplay between kinetic and interaction energy; this is in stark contrast to harmonically trapped gases, where the lowest mode frequency is independent of interaction strength \cite{kohn1961cyclotron}.

In this paper we experimentally and theoretically study the effect of interactions on the lowest-lying axial mode of a~\textsuperscript{87}Rb Bose-Einstein condensate (BEC) confined to a cylindrical box trap.
This mode was previously exploited as a route to turbulence in a continuously driven Bose gas~\cite{Turbulence}.
Here, we vary the atomic density by over two orders of magnitude to probe the near-equilibrium dynamics in both kinetic- and interaction-dominated regimes, and model our system using Bogoliubov theory to show how the mode evolves from a single-particle excitation to a sound wave.
We conclude by probing the response of the mode beyond the linear regime, revealing an intriguing non-exponential decay.

\section{Resonant Frequency}

\begin{figure}[b]
	\centering
	\includegraphics[width=\columnwidth]{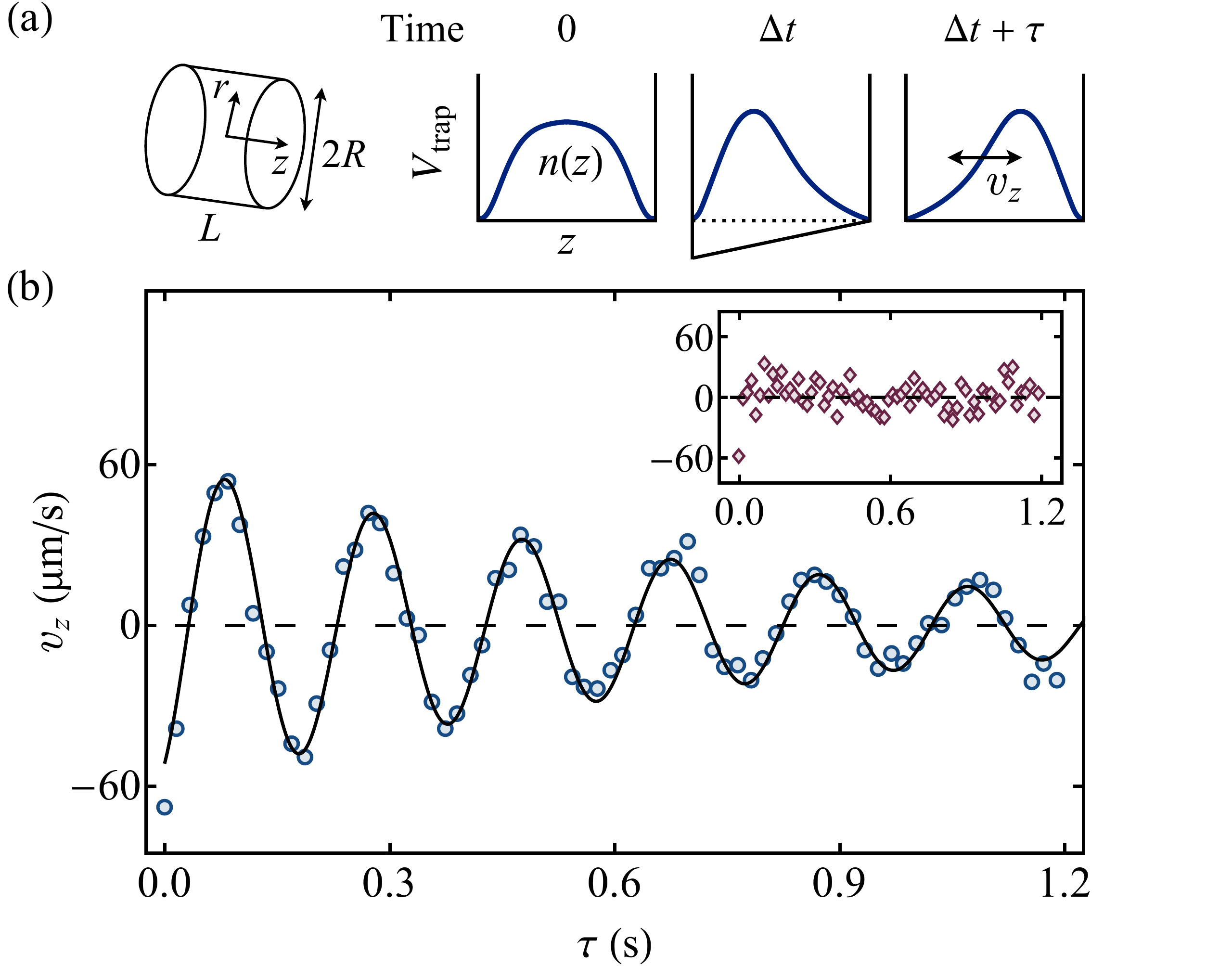}
		\caption{Probing the lowest axial mode. (a) Illustration of the experimental protocol. We prepare a BEC of \textsuperscript{87}Rb in a cylindrical box trap of length $L$ and radius $R$. Initially, the trapping potential $V_{\text{trap}}$ has a flat bottom. We then pulse a magnetic field gradient corresponding to $\Delta U=\kB \times 1$~nK along the box length for $\Delta t = 20$~ms. After a time $\tau$ of in-trap evolution we switch off the trap and let the cloud evolve in free space for $140$~ms before extracting its centre-of-mass, which reflects the velocity on release from the trap, $v_z$. (b) $v_z(\tau)$ for a BEC of $N = 13(1) \times 10^3$ atoms. We determine the oscillation frequency $\omega$ using a decaying sinusoidal fit. Inset: $v_z(\tau)$ for a non-condensed sample just above $T_{\textrm{c}}$, with $N=10(1) \times 10^3$.}
	\label{fig:experiment}
\end{figure}

Our experiments start with the production of quasi-pure BECs of between $N = 0.9 \times 10^3$ and $137 \times 10^3$ \textsuperscript{87}Rb atoms confined to a cylindrical optical box of length \mbox{$L=26(1)$~{\textmu}m} and radius \mbox{$R=16(1)$~{\textmu}m} (for details, see \cite{GauntUniform}).
The experimental protocol used to probe the axial mode is shown schematically in Fig.~\ref{fig:experiment}(a).
After creating the BEC we pulse an axial magnetic field gradient, corresponding to a potential difference $\Delta U = \kB \times 1$~nK over the box length, for a time \mbox{$\Delta t = 20$~ms}, short compared to the period of the mode.
We then hold the excited cloud in-trap for a variable time $\tau$ before switching off the trap and extracting the cloud's centre-of-mass (CoM) velocity in time-of-flight.
We observe an oscillation of the cloud's velocity with $\tau$, as shown in Fig.~\ref{fig:experiment}(b) for $N = 13(1) \times 10^3$ atoms.

If we repeat the same kick protocol with a thermal gas just above the condensation temperature, $T_{\textrm{c}}$, we see the same initial velocity as for a quasi-pure BEC. However, as shown in the inset of Fig.~\ref{fig:experiment}(b), there is no subsequent collective oscillation. For a classical gas to support hydrodynamic sound waves, local thermodynamic equilibrium needs to be established on timescales much shorter than the period of the wave~\cite{chaikin1995principles}. With wavelength $2L$ and speed $\sim \sqrt{k_\text{B} T/m}$, where $m$ is the atomic mass, this condition is equivalent to $L \gg 1/ (n a^2)$; the box length must be much greater than the mean free path.
Here $a$ is the $s$-wave scattering length (\mbox{$a \approx 100a_0$} for \textsuperscript{87}Rb, where $a_0$ is the Bohr radius), $n = N/V$ the atomic number density and $V = \pi R^2 L$ the volume of the trap. For our thermal gases this condition would be fulfilled only for $N \gg 10^7$ atoms, so hydrodynamic sound waves cannot propagate even in our densest samples. The oscillations we observe in the condensed gas at high density correspond to Bogoliubov sound waves.

\begin{figure}
	\includegraphics[width=\columnwidth]{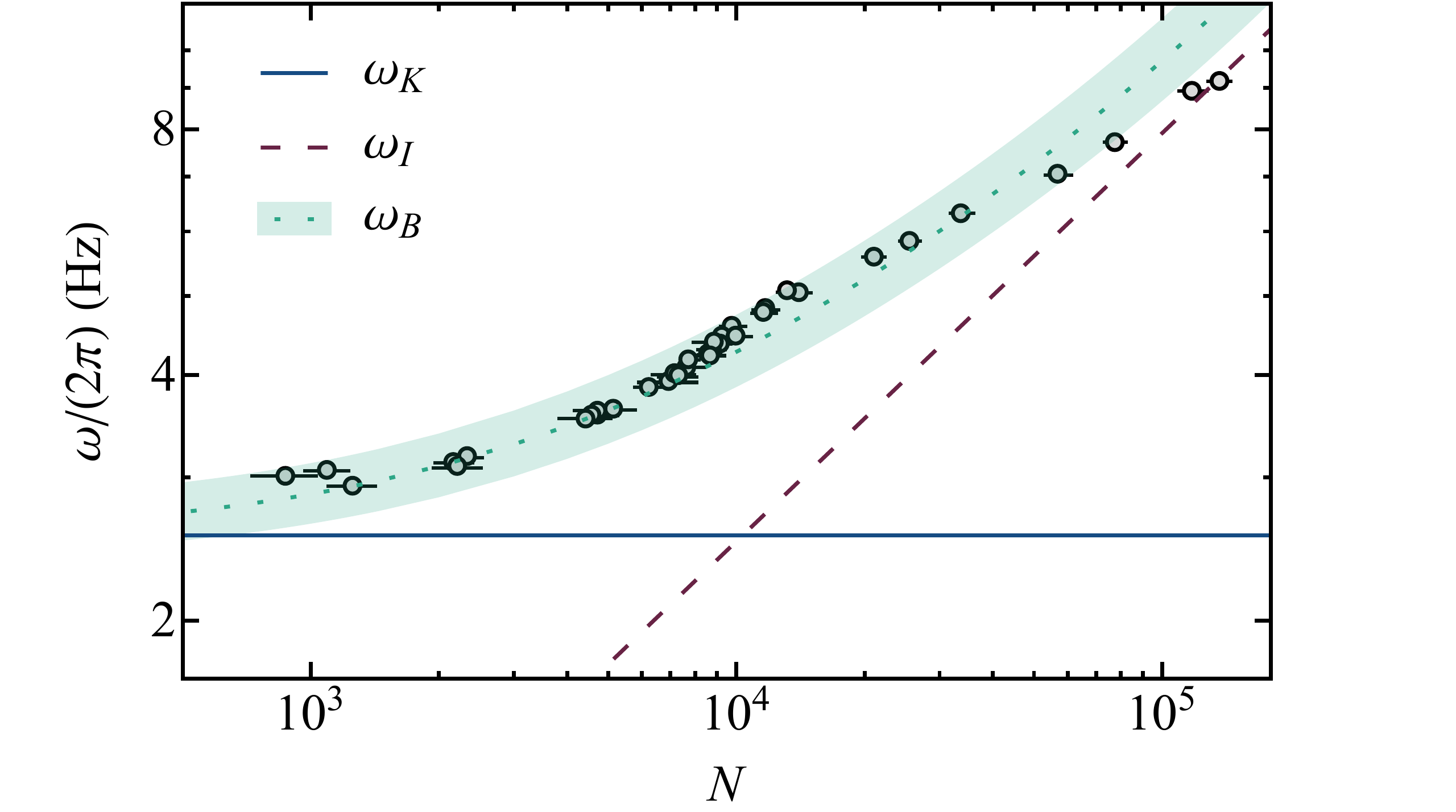}
	\caption{Angular frequency of the lowest axial mode as a function of atom number, $N$. $\omega_K$ is the mode frequency in an ideal gas, and $\omega_I$ is the frequency of Bogoliubov sound with speed $(g n/m)^{1/2}$ and wavelength $2L$. These calculations use trap dimensions $L = 26$~{\textmu}m and $R = 16$~{\textmu}m. The shaded band $\omega_B$ shows the results of numerical Bogoliubov diagonalisation (see Section \ref{ee}) accounting for the uncertainty of $\pm 1$~{\textmu}m in $L$ and $R$.}
	\label{fig:hydrodynamic}
\end{figure}

In Fig.~\ref{fig:hydrodynamic} we summarise the measured condensate oscillation frequencies, and compare them with theories in different interaction regimes. Throughout the paper we model the trap as an infinitely deep cylindrical potential well.

At low density the gas is kinetic energy-dominated and we expect ideal gas behaviour.
The system is then naturally described in terms of single-particle eigenstates $\alpha_j$, which are separable in cylindrical coordinates $(z, r, \phi)$.
The BEC wavefunction is simply the single-particle ground state $\alpha_0 = \varphi(r) \cos ( \pi z/L)$, with $\varphi(R)=0$, and the lowest axial mode corresponds to $\alpha_1 = \varphi(r) \sin (2 \pi z /L)$. The magnetic field gradient appears in the Hamiltonian as a perturbation $\hat H_{\textrm{kick}} = (\Delta U/L) \hat z$ for time $\Delta t$, and this preferentially excites particles from the condensate into $\alpha_1$. The excited state is then a superposition of $\alpha_0$ and $\alpha_1$, which exhibits velocity oscillations at angular frequency
\begin{equation}
    \omega_K = \frac{1}{\hbar}(\varepsilon_1 - \varepsilon_0) = \frac{3 \hbar}{2m} \Big( \frac{\pi}{L} \Big)^2, 
\label{eq:omegaK}
\end{equation}
where $\varepsilon_0$ and $\varepsilon_1$ are the single-particle energies, and the subscript $K$ denotes the kinetic-dominated regime. For our trap \mbox{$\omega_K / (2 \pi) = 2.5(2)$~Hz}.

In the interaction-dominated regime the condensate wavefunction is uniform away from the walls, where it decays over the healing length $\xi = (8 \pi n a)^{-1/2} \ll L$. The lowest axial mode is then a standing sound wave with wavelength $2L$ and speed of sound $(g n/m)^{1/2}$ \cite{pitaevskii2016bose}, where $g = 4 \pi \hbar^2 a /m$ is the strength of contact interactions. Consequently, at high density the axial mode frequency is
\begin{equation}
\omega_I = \Big(\frac{gn}{m}\Big)^{1/2} \frac{\pi}{L}.
\label{eq:omegaI}
\end{equation}

Between these limiting regimes, we capture the crossover with a numerical solution to the Bogoliubov equations (see Section \ref{ee} and \cite{Turbulence,Note3}). In the next section we investigate the physics of this crossover.

\section{Crossover}
Here we show how the condensate wavefunction of the interacting Bose gas, and its sound-wave excitations, emerge from the single-particle eigenstates $\alpha_j$. In second-quantised form, the Hamiltonian is
\begin{equation}
	\hat H = \underbrace{\sum_j \varepsilon_j \hat a_j^{\dag} \hat a_j}_{\hat K} +
	 \underbrace{\frac{g}{2V} \sum_{ijkl} I^{\alpha}_{ijkl} a_i^{\dag} \hat a_j^{\dag} \hat a_k \hat a_l}_{\hat I},
\label{eq:alphaH}
\end{equation}
where $\hat K$ and $\hat I$ are the kinetic- and interaction-energy operators, respectively. $\hat a_j^{\dag}$ is the creation operator for single-particle eigenstate $\alpha_j$, and $I^{\alpha}_{ijkl}/V = \int \alpha_i^* \alpha_j^* \alpha_k \alpha_l \, d^3 r$, where $V$ has been introduced so that $I^{\alpha}_{ijkl}$ is dimensionless.

\subsection{Condensate}
The non-interacting many-body ground state is \mbox{$\ket{\textrm{GS}} = (\hat a^{\dag}_0)^N \ket{0} / \sqrt{N!}$} where $\ket{0}$ is the vacuum of particles. Treating interactions perturbatively, the leading correction to the ground state, $\ket{\delta \textrm{GS}}$, comes from the operators of the form $\hat a_j^{\dag} \hat a_0^{\dag} \hat a_0 \hat a_0$. At first order
\begin{equation}
	\ket{\delta \textrm{GS}} = \frac{g}{V} \sum_{j \neq 0} I^{\alpha}_{j000} \frac{(N-1)\sqrt{N}}{\varepsilon_0 - \varepsilon_j} \frac{\hat a_j^{\dag} \hat a_0}{\sqrt{N}} \ket{\textrm{GS}},
\label{eq:nonperturbative}
\end{equation}
where $\hat a_j^{\dag} \hat a_0 \ket{\textrm{GS}}/ \sqrt{N}$ is the many-body state with one atom excited to single-particle eigenstate $\alpha_j$, and \mbox{$(gI^{\alpha}_{j000}/V) (N-1)\sqrt{N}$} is the matrix element between this many-body state and $\ket{\textrm{GS}}$. The squared norm of the correction, $\braket{\delta \textrm{GS} | \delta \textrm{GS}}$, is of order unity for $g n \sqrt{N} = \hbar \omega_K$, at which point perturbation theory fails. Note that in our experiment $g n \sqrt{N} = \hbar \omega_K$ for as few as $10^3$ atoms.

In order to develop a physical intuition for the role of the $\hat a_j^{\dag} \hat a_0^{\dag} \hat a_0 \hat a_0$ operators, we contrast our system with a condensate with periodic boundary conditions. In the periodic case the condensate is spatially uniform regardless of interaction strength, and all $I^{\alpha}_{j000}$ (for $j\neq 0$) vanish because $\alpha_j$ are momentum eigenstates. However, in our case these anomalous operators couple the even-parity eigenstates, thereby changing the condensate shape.

To show this explicitly, we first write the interacting condensate wavefunction $\beta_0$ as a superposition of single-particle eigenstates
\begin{equation}
	\beta_i = \sum_j U_{ij}(N) \alpha_j,
\label{eq:U}
\end{equation}
where $U_{ij}$ is a unitary transformation with a parametric dependence on the interaction strength, in our case captured by $N$. The states $\beta_{i \neq 0}$, which are orthogonal to each other and to $\beta_0$, will be used in the next section to construct the elementary excitations, but we first focus on the condensate. Working in the $\beta$ basis
\begin{equation}
\hat H = \sum_{ij} \braket{\beta_i | \hat K | \beta_j} \hat b^{\dag}_i \hat b_j + \frac{g}{2V} \sum_{ijkl} I^{\beta}_{ijkl} \hat b_i^{\dag} \hat b_j^{\dag} \hat b_k \hat b_l,
\label{eq:betaH}
\end{equation}
where $\hat b_i^{\dag}$ is the creation operator for $\beta_i$, and $I^{\beta}_{ijkl}$ are the corresponding overlap integrals.
Using particle number conservation \mbox{$\hat b_0^{\dag} \hat b_0 = N - \sum_{i \neq 0} \hat b_i^{\dag} \hat b_i$} we find
\begin{align}
	\hat H &= \sum_i \Big( (\braket{\beta_i|\hat K|\beta_0} + gnI^{\beta}_{i000}) \hat b^{\dag}_i \hat b_0 + \textrm{h.c.} \Big) + \hat H_2,
\end{align}
where we have dropped terms proportional to the identity, $\hat H_2$ is 2\textsuperscript{nd} order (and higher) in $\hat b_{i \neq 0}$ operators, and h.c.~denotes the Hermitian conjugate. If the ground state of $\hat H$ has a large condensate fraction, $\braket{\hat b_0^{\dag} \hat b_0} \gg \braket{ \hat b_i^{\dag} \hat b_i}$ for $i \neq 0$, we must have
\begin{equation}
	\braket{\beta_i|\hat K|\beta_0} + gnI^{\beta}_{i000} = 0,
\label{eq:nearly GP}
\end{equation}
for all $i \neq 0$; as $N$ increases the condensate wavefunction follows a path in the space of single-particle eigenstates along which it is decoupled from all orthogonal states.

If the set of $\beta_i$ forms a complete basis, then Eq.~\eqref{eq:nearly GP} implies
\begin{equation}
\Big( -\frac{\hbar^2}{2m}\nabla^2 + g n |\beta_0|^2 \Big) \beta_0 = \mu \beta_0,
\label{eq:GP}
\end{equation}
for a constant $\mu$, which is identified as the Hartree-Fock chemical potential. Eq.~\eqref{eq:GP} is then the well-known Gross-Pitaevskii (GP) equation~\cite{pitaevskii2016bose}, which has arisen from the sole assumption that one state, $\beta_0$, has much greater occupation than those orthogonal to it.

In practice, we work with a truncated basis of single-particle eigenstates. Minimising the GP energy functional (with respect to $U_{0j}$) within this truncated set gives the condensate wavefunction $\beta_0$ that is not an exact solution to the GP equation, but does satisfy Eq.~\eqref{eq:nearly GP}.

\begin{figure}[t]
	\centering
	\includegraphics[width=\columnwidth]{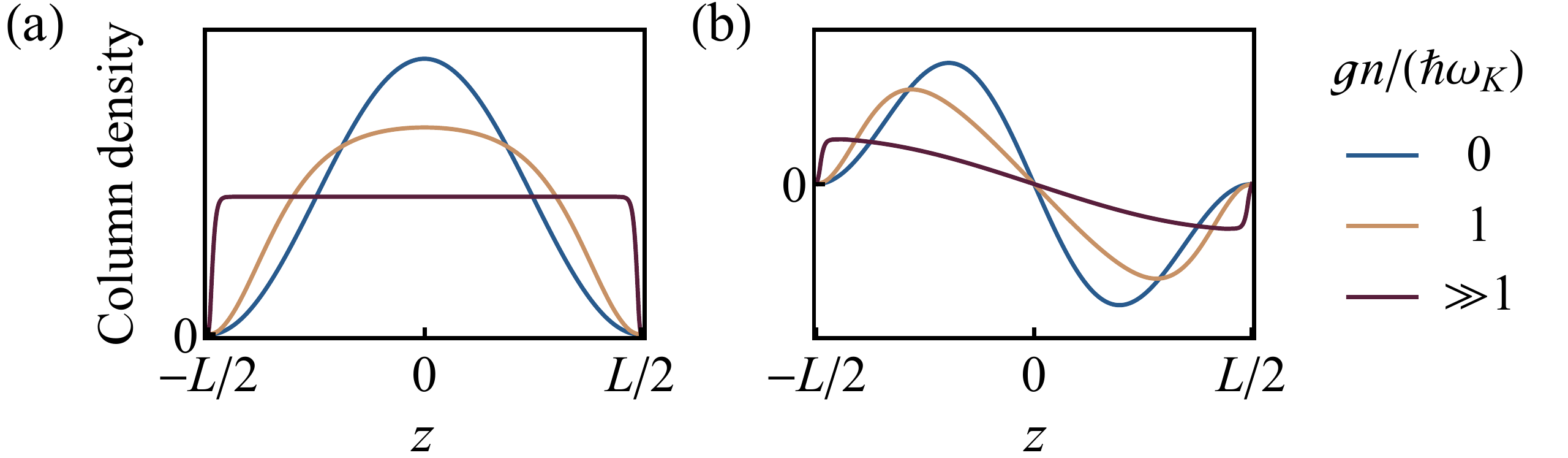}
	\caption{(a) Axial density profile of the BEC, for three interaction strengths $gn/(\hbar \omega_K)$. (b) Change in the axial density profile due to the coherently-occupied axial mode, scaled by $\sqrt{\omega}$ so the amplitude for $gn/(\hbar \omega_K) \gg 1$ does not depend on $N$.}
	\label{fig:shapes}
\end{figure}

In Fig.~\ref{fig:shapes}(a) we illustrate how the condensate shape changes in the crossover between kinetic- and interaction-energy dominated regimes.

\subsection{Excitations}\label{ee}

To study the evolution of excitations with interaction strength we use Bogoliubov theory. Having determined $\beta_0$ variationally, we construct the set of $\beta_{i \neq 0}$ using the Gram-Schmidt procedure, then introduce the mean field $\hat b_0 = \sqrt{N}$ in Eq.~\eqref{eq:betaH}. As in conventional Bogoliubov theory, we neglect terms cubic (and higher) in $\hat b_{i \neq 0}$ operators, thereby arriving at an effective quadratic Hamiltonian for the near-equilibrium dynamics. This is diagonalised using a bosonic transformation~\cite{blaizot1986quantum}
\begin{equation}
	\hat c_i = \sum_{j=1} \Big( P_{ij} \hat b_j + Q_{ij} \hat b^{\dag}_j \Big),
\label{eq:bosonictransform}
\end{equation}
where $P$ and $Q$ are chosen such that
\begin{equation}
	\hat H \approx \hat H_B = \sum_{j=1} \hbar \omega_j \hat c^{\dag}_j \hat c_j
\label{eq:BogoliubovH}
\end{equation}
is the Bogoliubov Hamiltonian, where we have omitted the energy of the interacting ground state and $\hat c_j^{\dag}$ is the creation operator for the $j^{\textrm{th}}$ normal mode. Although the Gram-Schmidt procedure does not uniquely specify $\beta_{i \neq 0}$, $P_{ij}$ and $Q_{ij}$ adjust accordingly to uniquely specify $\hat c_i$ (up to a phase factor). 

In Fig.~\ref{fig:shapes}(b) we show how the axial density profile of the lowest-lying axial mode changes through the crossover.
Note that the wavelength of the mode in the interaction-dominated regime is double of that in the ideal gas.

Before comparing our theoretical results with the experiments, it remains to be shown that throughout the crossover the axial kick $\hat H_{\text{kick}}$ leads to velocity oscillations at the frequency of the lowest axial mode. Here we show that this is indeed the case, and that the excited state has a coherent occupation of normal modes. Approximating $\hat b_0 = \sqrt{N}$ and assuming $N \gg 1$, we express $\hat H_{\textrm{kick}}$ as a linear combination of normal mode operators
\begin{align}
	\hat H_{\textrm{kick}} &\approx \frac{\Delta U}{L}\sqrt{N} \sum_{i} \braket{\beta_0|\hat z|\beta_i} \hat b_i + \textrm{h.c.}, \nonumber \\
	&= \frac{\Delta U}{L}\sqrt{N} \sum_j Z_{j} \hat c_j + \textrm{h.c.}, 
\label{eq:2nd_z}
\end{align}
where $Z_j$ is given by the $P$ and $Q$ matrices \footnote{Defining the inverse matrices $P^{-1}$ and $Q^{-1}$ via $\hat b_i = \sum_j [P^{-1}_{ij} \hat c_j + Q^{-1}_{ij} \hat c_j^{\dag}]$, we have $Z_j = \sum_i [ \braket{\beta_0| \hat z |\beta_j}P^{-1}_{ij} + \braket{\beta_j| \hat z |\beta_0} (Q^{-1}_{ij})^* ]$}. We then treat $\hat H_{\textrm{kick}}$ as a perturbation to $\hat H_B$ and find the time-evolution operator in the interaction picture,
\begin{equation}
\hat U(\Delta t) = \prod_{j} \mathcal{T} \exp \Big( - i \int_0^{\Delta t}  ( \eta_j e^{-i \omega_j t'} \hat c_j + \textrm{h.c.} ) dt' \Big), 
\label{eq:time evolution}
\end{equation}
where $\eta_j \equiv \Delta U \sqrt{N} Z_j / (\hbar L$) and $\mathcal{T}$ denotes time ordering. For the low energy modes $e^{-i \omega_j \Delta t}~\approx~1$, so the time ordering operation is trivial.
To obtain the state at the end of the kick we apply $\hat U(\Delta t)$ to the ground state of $\hat H_B$, $\ket{\textrm{GS}}_B$, which yields
\begin{equation}
\ket{\eta} \equiv \hat U(\Delta t)\ket{\textrm{GS}}_B = \prod_{j} e^{-|\eta_j \Delta t|^2 /2} e^{ - i \eta_j \Delta t \hat c^{\dag}_j} \ket{\textrm{GS}}_B,
\label{eq:eta}
\end{equation}
a coherent occupation of normal modes.
Following the kick, the in-trap time evolution is generated by $\hat H_B$.
To calculate the velocity of the cloud we first write the momentum operator in terms of the normal modes. Following the same procedure as in Eq.~\eqref{eq:2nd_z},
\begin{equation}
	\hat p_z \approx \hbar \sqrt{N} \sum_j D_j \hat c_j + \textrm{h.c.},
\end{equation}
where $\hbar D_j$ is analogous to $Z_j$ \footnote{Using the inverse matrices $P^{-1}$ and $Q^{-1}$, we have $\hbar D_j = \sum_i [ \braket{\beta_0| \hat p_z |\beta_j}P^{-1}_{ij} + \braket{\beta_j| \hat p_z |\beta_0} (Q^{-1}_{ij})^* ]$}.
The axial velocity is then
\begin{align}
	\braket{v_z(\tau)} &= \frac{1}{Nm} \bra{\eta} e^{i \hat H_B \tau /\hbar} \hat p_z e^{-i \hat H_B \tau/\hbar}\ket{\eta} \nonumber \\
	&= 2\frac{\Delta t}{m}\frac{\Delta U}{L} \sum_j \textrm{Im} \Big\{ D_j Z_j e^{- i \omega_j \tau} \Big\},
\end{align}
The contribution of the $j^{\textrm{th}}$ mode to the velocity amplitude therefore scales as $|Z_j D_j|$. We find numerically that the lowest axial mode contributes 96\% of the amplitude in our most dilute samples, falling to 86\% in our densest ones. Moreover, as the excitation spectrum is discrete, any significant population of higher modes would cause the velocity oscillations to be visibly non-sinusoidal, which we do not observe experimentally. All of this confirms that the oscillations that we observe (see Fig.~\ref{fig:experiment}(b)) arise as a result of the direct coupling to the lowest-axial mode.

\begin{figure}[b]
	\centering
	\includegraphics[width=\columnwidth]{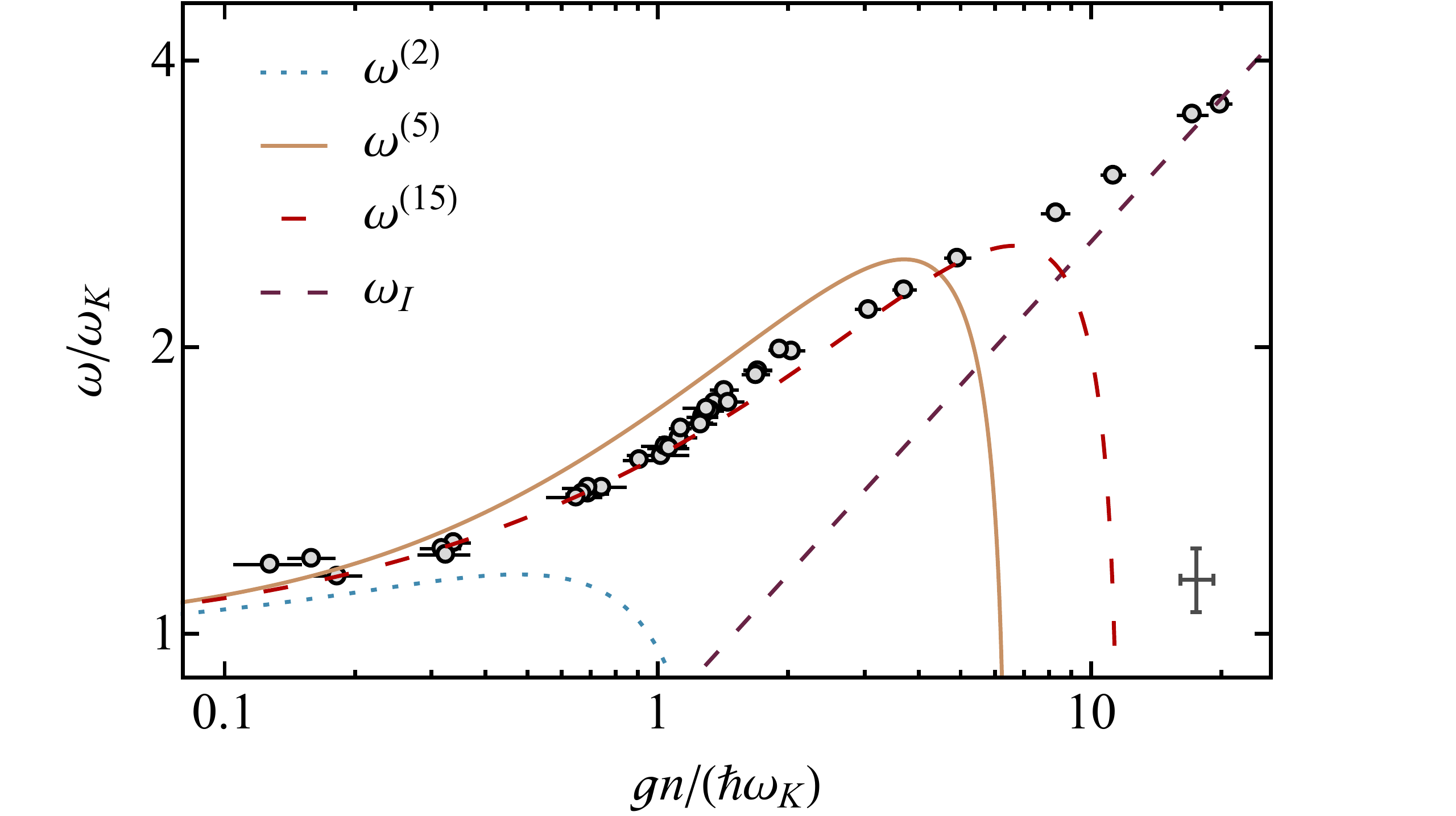}
	\caption{
	Crossover from single-particle excitations to sound~waves.
	We plot $\omega/\omega_K$ as a function of the interaction strength, now parametrised by $gn / (\hbar \omega_K)$.
	Here we assume $L = 26~${\textmu}m and $R = 16~${\textmu}m for calculations. The $x$-$y$ error bar in the bottom right corner indicates the fractional systematic uncertainties in the experimental data due to the uncertainties in the box dimensions.
	$\omega^{(2)}$ is determined using Bogoliubov theory within a truncated basis of just the two lowest-energy single-particle eigenstates of zero angular momentum (see text).
	This scheme fails even for relatively small $gn/(\hbar \omega_K)$, as it does not allow for the interaction-induced changes in the shape of the condensate or the excitation mode.
	$\omega^{(5)}$, based on a truncated basis of five single-particle eigenstates, is the minimal model that allows for the shape changes.
	This simple model already captures most of the crossover, and using progressively larger truncated bases does not qualitatively change the result, as shown by $\omega^{(15)}$, which is based on $15$ single-particle eigenstates. $\omega_I$ is the sound-wave frequency, approached in the limit of large $gn/(\hbar \omega_K)$.}
	\label{fig:interaction_shift}
\end{figure}

\subsection{Truncated-basis models}\label{truncated}

To understand the effect of interactions on the axial mode frequency throughout the crossover, we calculate it using (progressively larger) truncated sets of low-energy single-particle eigenstates.

First, we consider just the two lowest single-particle eigenstates (the lowest even-parity state $\alpha_0$ and the lowest odd-parity state $\alpha_1$). In this case the condensate wavefunction is $\beta_0 = \alpha_0$, the mode involves only $\alpha_1$, and neither has the freedom to change its shape with increasing $g n / (\hbar \omega_K)$.
Here the bosonic transform in Eq.~\eqref{eq:bosonictransform} gives $\hat c_1 = \cosh(\kappa) \hat a_1 + \sinh(\kappa) \hat a_1^{\dag}$, where $\kappa$ is chosen to diagonalise $\hat H_B$ (see Eq.~\eqref{eq:BogoliubovH}). The resulting mode frequency is
\begin{equation}
	\omega^{(2)} = \omega_K \sqrt{ \Big{(}1 + \frac{\mathcal{J} gn}{3\hbar \omega_K}\Big{)}^2 - \Big{(}\frac{2 \mathcal{J} gn}{3\hbar \omega_K}\Big{)}^2 },
\end{equation}
where $\mathcal{J} \approx 2.10$ is the radial factor in the overlap integral $I^{\alpha}_{0011}$. As shown in Fig.~\ref{fig:interaction_shift}, this scheme fails to describe the dynamics even for values of $g n / (\hbar \omega_K)$ well below unity.

As a minimal model that does allow for the interaction-induced shape changes of both the condensate and the excitation mode, we consider a truncated set of the five lowest-energy $\alpha_j$ with zero angular momentum, and calculate the corresponding $\omega^{(5)}$. In this case $\beta_0$ is a superposition of three even-parity states, and the excited mode involves two of odd-parity; note that here we use five states because the third-lowest even state has a lower energy than the second-lowest odd one.
We find that this simple model is sufficient to capture rather well most of the crossover, up to $gn/(\hbar \omega_K)\approx 3$ (see Fig.~\ref{fig:interaction_shift}), where $gn$ exceeds the maximum kinetic energy in the truncated set.
For any $gn/(\hbar \omega_K)$ significantly larger than $3$, the simple sound-wave calculation $\omega_I$ already provides a good approximation.

The agreement of the Bogoliubov calculations with the experimental data improves further as we consider ever larger basis sets, as shown by the 15-state calculation, $\omega^{(15)}$, in Fig.~\ref{fig:interaction_shift}, and the full numerical result in Fig.~\ref{fig:hydrodynamic}; there $\omega_B$ was calculated using $80$ single-particle states with kinetic energies up to $\approx 70~\hbar \omega_K$.
However, the success of the simple $\omega^{(5)}$ calculation highlights the key qualitative message: most of the physics of the crossover is captured by including, at lowest order, the interaction-induced shape changes that arise due to the experimentally relevant fixed boundary conditions. 

\section{Beyond Linear Response}

We have so far neglected any coupling between the normal modes of the BEC. However, we observe at least a weak damping in all the frequency measurements summarised in Figs.~\ref{fig:hydrodynamic} and  \ref{fig:interaction_shift}; see for instance Fig.~\ref{fig:experiment}(b). In general, the nonzero temperature of our gases will lead to Landau damping~\cite{pitaevskii1997landau}, but our conservative upper bound of $T < 10 \, \textrm{nK}$ suggests a decay rate of \mbox{$\Gamma_{\textrm{Landau}}/(2\pi) < 0.2~$s$^{-1}$}, much smaller than observed.
\begin{figure}
	\centering
	\includegraphics[width=\columnwidth]{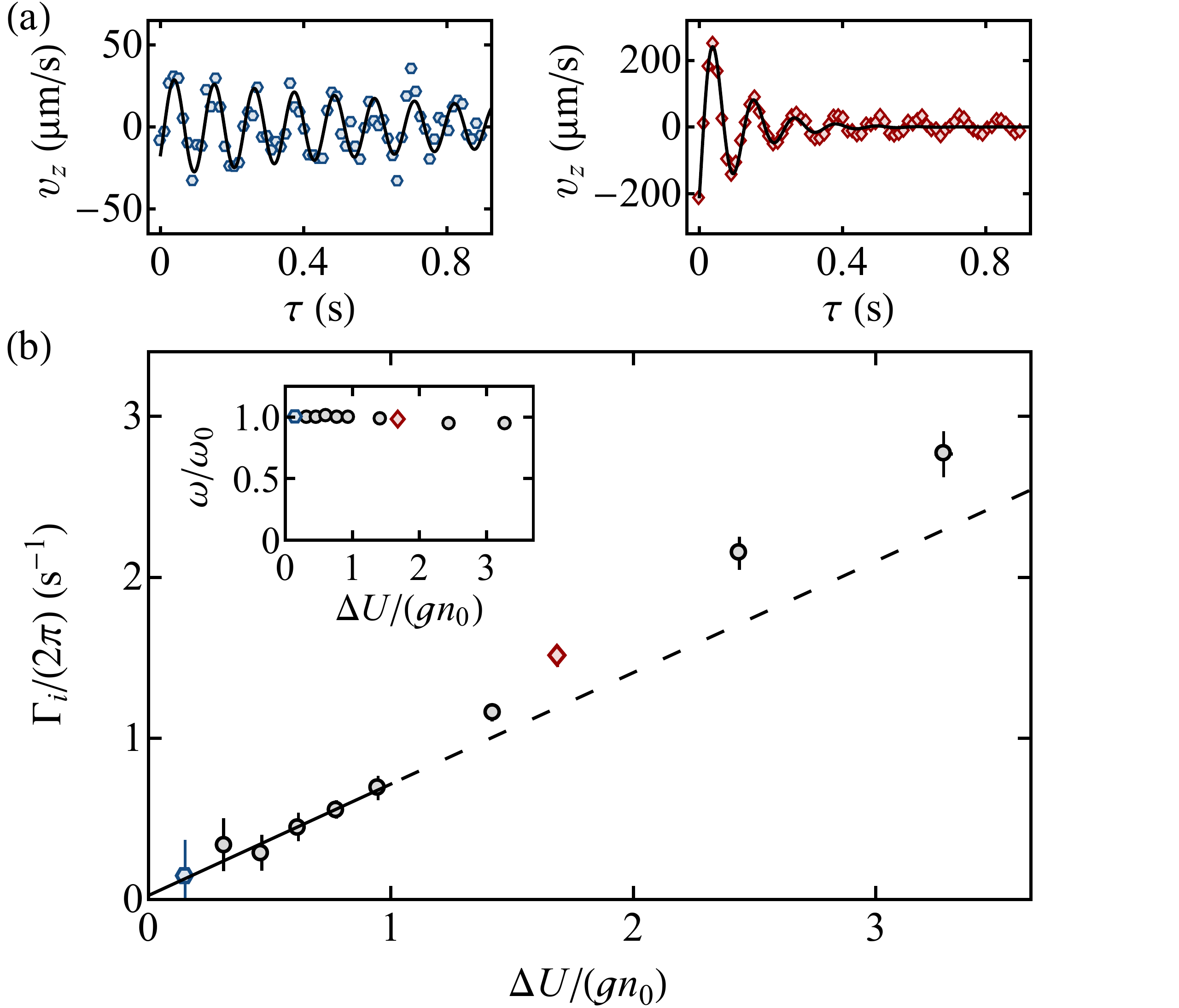}
	\caption{Nonlinear damping of the lowest-lying axial mode in the interaction-dominated regime, with fixed density $n_0$ such that $gn_0 = k_\text{B} \times 2.1$~nK (corresponding to $gn_0 /(\hbar \omega_K) \approx 17$). (a) Velocity oscillations following kicks with $\Delta U= k_\text{B}\times 0.3$~nK (left) and $k_\text{B} \times 3.6$~nK (right). The solid lines show exponentially decaying sinusoidal functions obtained by fitting to the early-time data, $\tau<0.25$~s. For large $\Delta U$ (right panel) the decay is clearly non-exponential. (b) Initial decay rate $\Gamma_i$ as a function of the normalised kick amplitude $\Delta U/(gn_0)$. The solid line is a linear fit to data with $\Delta U/(gn_0)<1$ (the dashed line displays its extrapolation for $\Delta U/(gn_0)>1$). Inset: frequency of the damped oscillation $\omega$ normalised to the low-$\Delta U$ value, $\omega_0$.}
	\label{fig:damping}
\end{figure}

Here we examine this damping for different kick amplitudes $\Delta U$, focusing on the interaction-dominated regime with $gn/(\hbar \omega_K) \approx 17$; we fix $gn$ to $gn_0 = k_\text{B} \times 2.1(2)$~nK by fixing the atom number to $N = 1.2(1) \times 10^5$.
Fig.~\ref{fig:damping}(a) shows $v_z(\tau)$ following kicks with $\Delta U= k_\text{B}\times 0.3$~nK (left panel) and $k_\text{B} \times 3.6$~nK (right panel).
For the weak kick only a subtle damping is observed, and an exponentially decaying sine (solid line), fit to the early-time data ($\tau<0.25$~s), captures the data well for all $\tau$. However, for the stronger kick we clearly see a rapid initial decay followed by a long-lived oscillation. Here the solid line, based on the same fitting to the early-time data ($\tau<0.25$~s), clearly fails to capture the oscillations for $\tau \gtrsim 0.4$~s.

We characterise the damping using the initial velocity-decay rate, $\Gamma_i$, extracted from the early-time ($\tau <0.25$~s) fits.
In Fig.~\ref{fig:damping}(b) we show $\Gamma_i$ versus normalised kick amplitude, $\Delta U/(gn_0)$, and in the inset we show that the mode frequency is approximately constant across our whole range of $\Delta U$. For relatively weak kicks ($\Delta U \lesssim gn_0$) the damping rate appears to be linear in kick amplitude, essentially vanishing (within experimental errors) as $\Delta U \rightarrow 0$. This diverging lifetime in the limit of vanishing excitation amplitude is consistent with the absence of lower-lying modes to which this mode could readily decay, and moreover it excludes (at the level of our experimental errors) damping due to non-zero temperature or technical reasons. Both the non-exponential decay and the fact that $\Gamma_i$ increases with $\Delta U$  suggest that the damping occurs due to interactions between the excitations.

\section{Conclusions and outlook}
We have measured the dynamics of an atomic BEC in a cylindrical box trap following an axial kick, thereby probing its lowest axial mode. By tuning the gas density we studied the evolution from single-particle to many-body dynamics. We used a simple model to elucidate the effect of interactions, and numerically evaluated the mode frequency over the whole range of densities, finding excellent agreement with the experiments. Going beyond linear response in the interaction-dominated regime, we observed a non-exponential decay of the excitation, hinting at a nonlinear many-body decay mechanism. A future challenge is to understand this decay mechanism, and in particular its dependence on the interaction strength.

\section{Acknowledgements}
We are grateful to F. Werner and J. T. Chalker for comments on the manuscript. This work was supported by EPSRC [Grants No.~EP/N011759/1 and No.~EP/P009565/1], ERC (QBox), AFOSR, and ARO. N.N. acknowledges support from Trinity College (Cambridge) and the David and Lucile Packard Foundation. R.L. acknowledges support from the E.U. Marie-Curie program [Grant No.~MSCA-IF-2015 704832] and Churchill College, Cambridge.  R.P.S. acknowledges support from the Royal Society. \\


%


\end{document}